\documentclass[twocolumn,prl,showpacs,preprintnumbers,amsmath,amssymb]{revtex4}
\usepackage{epsf,amsmath,amssymb,verbatim,color,multirow,pifont}
\usepackage{graphicx}

\begin{document}
\title{Segregation in the annihilation of two-species reaction-diffusion processes 
on fractal scale-free networks}
\author{C.-K. Yun}
\author{B. Kahng}
\author{D. Kim}
\affiliation{Department of Physics and Astronomy and Center for
Theoretical Physics, Seoul National University, Seoul 151-747, Korea}
\date{\today}
\begin{abstract}
In the reaction-diffusion process $A+B \to \varnothing$ on random scale-free (SF) networks
with the degree exponent $\gamma$, the particle density decays with time in a power law
with an exponent $\alpha$ when initial densities of each species are the same.
The exponent $\alpha$ is $\alpha > 1$ for $2 < \gamma < 3$
and $\alpha=1$ for $\gamma \ge 3$. Here, we examine the reaction process on fractal SF networks, finding that $\alpha < 1$ even for $2 < \gamma < 3$. This slowly decaying behavior originates from the segregation effect: Fractal SF networks contain local hubs, which are repulsive to each other. Those hubs attract particles and accelerate the reaction, and then create domains containing the same species of particles. It follows that the reaction takes place at the non-hub boundaries between those domains and thus the particle density decays slowly. Since many real SF networks are fractal, the segregation effect has to be taken into account
in the reaction kinetics among heterogeneous particles.
\end{abstract}
\maketitle

Diffusion-limited reaction kinetics has been studied for long time as an inter-disciplinary subject. It can be a model of electron-hole recombination in semiconductors~\cite{e_hole} and annihilation of primordial monopoles in the early universe~\cite{Toussaint,klee}, etc. The annihilation process involving two species $A$ and $B$ of particles $A+B \to \varnothing$ is studied here, particularly on fractal SF networks. When the densities of $A$ and $B$ particles are initially equal, the density of each species  $\rho_A(t)$ or $\rho_B(t)$ decays in a power law, that is, $\rho_A(t)=\rho_B(t)\equiv \rho(t)\sim t^{-\alpha}$. In a mean-field approximation, the density of particles decays as $\rho(t)\sim t^{-1}$, which is valid when the reaction takes place in Euclidean space with the spatial dimension $d > d_c=4$. For $d < d_c$, the exponent $\alpha$ is reduced to $\alpha=d/d_c$, which is less than 1. The slow decaying behavior of particle density  originates from the formation of $A$-rich or $B$-rich domains, and the reaction takes place at the boundary of those domains in Euclidean space~\cite{kiho,leyvraz,kopelman}.

In complex networks, however, the particle density $\rho(t)$ can decay faster than the mean-field behavior $\rho(t)\sim t^{-1}$ in the long time limit~\cite{gallos2004}.
This fast decay is caused by the existence of hubs at which particles gather through the diffusion process and then reaction takes place frequently. It is noteworthy that the probability of finding a random walker at a node is proportional to the degree of that node~\cite{sneppen,jdnoh}.
For the uncorrelated SF networks, the particle density $\rho(t)$ was derived analytically
for $A+A \to \varnothing$~\cite{catanzaro} and
$A+B \to \varnothing$~\cite{weber2006} as
\begin{equation}
\frac{1}{\rho(t)}-\frac{1}{\rho(0)}=
\begin{cases}
t^{1/(\gamma-2)} & \text{for}~2< \gamma < 3,\\
t\ln t & \text{for}~\gamma=3, \\
t & \text{for}~\gamma >3,
\end{cases}
\label{rt_origin} \end{equation}
where $\gamma$ is the exponent of the degree distribution $P_{d}(k)\sim k^{-\gamma}$ of the SF networks.

In this Letter, we demonstrate that when SF networks are fractal~\cite{song,goh},
the segregation of $A$-rich or $B$-rich domains arises, and the particle density decays
slowly with the exponent $\alpha \le 1$, different from the formula (\ref{rt_origin}).
Fractal SF network is a network satisfying the fractal scaling $N_{B}(\ell_{B})\sim \ell_B^{-d_f}$, where $N_{B}$ is the number of boxes needed to cover the entire network with boxes of size $\ell_{B}$. The fractal scaling holds when hubs are located separately from others in position~\cite{yook,song2}. Many SF network observed in real world are fractals. Note that most artificial networks including Barab\'sasi and Albert (BA) model~\cite{ba} are not fractals~\cite{jskim}. In the fractal networks, local hubs attract particles and accelerate the reaction. As a result, in early time regime, particle density decreases rapidly with $\alpha > 1$. After that period, domains are created in which the same species of particles remain, which are the majority induced by fluctuations of initial particle densities. Then, the reaction takes place only at the boundary between those domains, which are not hubs. Thus the particle density decays slowly in the long time limit with $\alpha < 1$. Such segregation behavior can also occur in modular SF networks, even if they are non-fractals. Structural feature of the modular network, being composed of a large number of links within modules but a small number of links between modules, hampers the diffusion of particles across modules.

To study the two-species reaction $A+B \to \varnothing$ on fractal SF networks specifically, we first recall the previous studies~\cite{leyvraz,kopelman} of the reaction kinetics taking place on fractal structure embedded in Euclidean space. In this case, the formula $\rho(t)\sim t^{-d/d_c}$ may be replaced with
\begin{equation}
\rho(t)\sim t^{-d_s/4},
\label{rt_spec}
\end{equation}
where $d_s$ is the spectral dimension of the fractal structure. $d_s$ is related to
random walk dimension $d_w$ and fractal dimension $d_f$ as $d_s=2d_f/d_w$.
The random walk dimension is defined through the anomalous power-law relationship between the mean-square displacement $\langle \ell^2(t) \rangle$ of a diffusing particle and time $t$ as $\langle \ell^2(t) \rangle \sim t^{2/d_w}$.
The formula (\ref{rt_spec}) has been questioned, however, because it does not take into account of structural features in a given fractal structure such as the degree of ramification. Nevertheless, it appears that numerical results are essentially in agreement with this prediction (\ref{rt_spec}) for many cases~\cite{meakin,zumofen}. In this Letter, we show that in contrast to the standard random SF network cases,  for the fractal SF networks we study here, the particle density decays in the form given by (\ref{rt_spec}).

Here we first generate a fractal SF tree structure through the multiplicative branching process. At each branching step, a node creates its $m$ branches (offsprings) with probability $p_m\sim m^{-\gamma}$ ($m\ge 1$). It has to satisfy the criticality condition $\langle m \rangle=\sum_{m=0}^{\infty} mp_m=1$~\cite{goh}. Then, the resulting tree structure is a SF tree with the degree exponent $\gamma$. Such a random critical branching tree structure is a fractal SF network with the fractal dimension $d_f=(\gamma-1)/(\gamma-2)$ for $2 < \gamma < 3$ and $d_f=2$ for $\gamma > 3$. The spectral dimension is $d_s=2(\gamma-1)/(2\gamma-3)$ for $2 < \gamma < 3$ and $d_s=4/3$ for $\gamma > 3$~\cite{burda,makse}.

\begin{figure}[h]
\includegraphics[width=5.5cm, angle=-90]{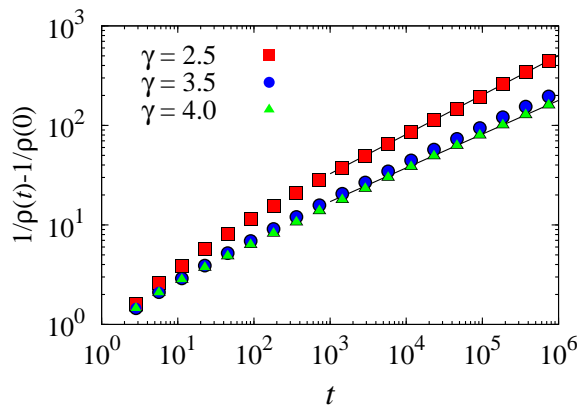}
\caption{(Color on line) The particle density as a function of time on the critical
branching trees with various degree exponents. Guidelines have slopes $0.40$ (top) and $0.34$ (bottom).}\label{ft_rho}
\end{figure}

We measure particle density $\rho(t)$ as a function of time $t$ in the form,
\begin{equation}
\frac{1}{\rho(t)}-\frac{1}{\rho(0)}\sim t^{\alpha}.
\end{equation}
We find that the particle density decays fast in short time regime,
followed by a slow decay in the long time regime as shown in Fig.\ref{ft_rho}. Indeed, numerically obtained values listed in Table~\ref{table1} are close to the one obtained from the formula $d_s/4$, and different from the ones obtained from the formula~(\ref{rt_origin}).

\begin{table}[h]
\caption{Comparison of the exponent $\alpha$ numerically obtained, denoted as $\alpha_{\rm num}$, with $d_s/4$ for various degree exponent $\gamma$'s
for the critical branching tree. For comparison, we also
list the mean field value obtained from the formula (\ref{rt_origin}).}
\begin{center}
\begin{tabular}{llll}
\hline\hline
~~~~$\gamma$~~~~ & ~~~~$\alpha_{\rm num}$~~~~ & ~~~~$d_s/4$~~~~ & ~~~~{\rm MF value}~~~~ \\
\hline
~~~~2.5~~~~ & ~~~~0.4~~~~  & ~~~~0.38~~~~ & ~~~~~~~~2.00~~~\\
~~~~2.7~~~~ & ~~~~0.36~~~  & ~~~~0.35~~~~ & ~~~~~~~~1.43~~~\\
~~~~3.5~~~~ & ~~~~0.35~~~  & ~~~~0.33~~~~ & ~~~~~~~~1.00~~~\\
~~~~4.0~~~~ & ~~~~0.34~~~  & ~~~~0.33~~~~ & ~~~~~~~~1.00~~~\\
~~~~4.5~~~~ & ~~~~0.34~~~  & ~~~~0.33~~~~ & ~~~~~~~~1.00~~~\\
\hline
\end{tabular}
\end{center}
\label{table1}
\end{table}

Next, we study the reaction kinetics on deterministic fractal
SF networks, introduced by Rozenfeld et al.~\cite{rozenfeld}, the so called
($u,v$)-flower and ($u,v$)-tree networks. These networks are hierarchical networks,
generated iteratively from a simple basic structure to higher level ones.
Each link in the $n$-th generation is replaced by two parallel paths of $u$
and $v$ links long. Detailed rule can be found in Ref.~\cite{rozenfeld}.
Depending on the rule, constructed networks are either the flower structure
which contains loops or trees. These networks are fractal SF networks with
the degree exponent,
$\gamma=1+\frac{\ln(u+v)}{\ln 2},$
the fractal dimension, $d_f=\frac{\ln(u+v)}{\ln u}$,
and the spectral dimension,
$d_s=\frac{2\ln(u+v)}{\ln uv}$ for flowers, and $\frac{2\ln(u+v)}{\ln u(u+v)}$ for trees.
Numerical values of the exponent $\alpha$ are close to those from $\alpha=d_s/4$
as can be seen in Tables II and III for the flower and tree structures, respectively.
\begin{figure}[h]
\includegraphics[width=5.5cm, angle=-90]{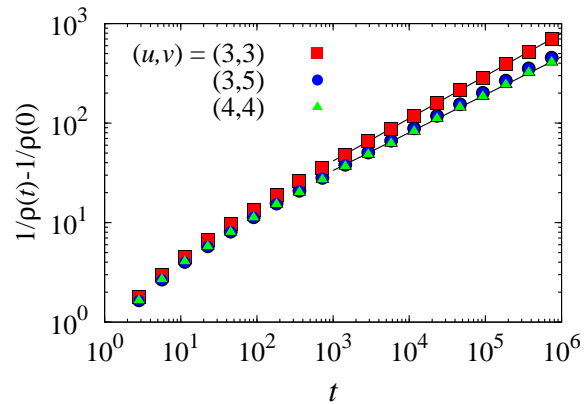}
\caption{(Color on line) The particle density of $A$ or $B$ species as a function
of time for the ($u,v$)-flower networks. Guidelines have slopes $0.43$ (top) and $0.38$ (bottom).}\label{uv-flower}
\end{figure}

\begin{table}[h]
\caption{Comparison of the exponent $\alpha$ numerically obtained,
denoted as $\alpha_{\rm num}$, with $d_s/4$ for various
degree exponent $\gamma$'s for the ($u,v$)-flower networks. For comparison, we also
list the mean field value obtained from (\ref{rt_origin}).}
\begin{center}
\begin{tabular}{lllll}
\hline\hline
~~~$(u,v)$~~~~&~~~~~$\gamma$~~~~&~~~$\alpha_{\rm num}$~~~&~~~$d_s/4$~~~&~~{\rm MF value}~~~\\
\hline
~~~(2,2)~~~&~~~3~~~&~~~0.53~~~&~~~~0.5~~~&~~1.0~~~\\
~~~(2,4)~~~&~~~3.58~~~&~~~0.45~~~&~~~~0.43~~~&~~1.0~~~\\
~~~(3,3)~~~&~~~3.58~~~&~~~0.43~~~&~~~~0.41~~~&~~1.0~~~\\
~~~(2,6)~~~&~~~4~~~&~~~0.43~~~&~~~~0.42~~~&~~1.0~~~\\
~~~(4,4)~~~&~~~4~~~&~~~0.38~~~&~~~~0.38~~~&~~1.0~~~\\
\hline
\end{tabular}
\end{center}
\label{table2}
\end{table}

\begin{table}[h]
\caption{The same as Table II for the ($u,v$)-tree networks.}
\begin{center}
\begin{tabular}{lllll}
\hline\hline
~~~$(u,v)$~~~~&~~~~~$\gamma$~~~~&~~~$\alpha_{\rm num}$~~~&~~~$d_s/4$~~~&~~{\rm MF value}~~~\\
\hline
~~~(2,2)~~~&~~~3~~~&~~~0.34~~~&~~~~0.33~~~&~~1.0~~~\\
~~~(2,4)~~~&~~~3.58~~~&~~~0.37~~~&~~~~0.36~~~&~~1.0~~~\\
~~~(3,3)~~~&~~~3.58~~~&~~~0.31~~~&~~~~0.31~~~&~~1.0~~~\\
~~~(2,6)~~~&~~~4~~~&~~~0.38~~~&~~~~0.38~~~&~~1.0~~~\\
~~~(4,4)~~~&~~~4~~~&~~~0.31~~~&~~~~0.30~~~&~~1.0~~~\\
\hline
\end{tabular}
\end{center}
\label{table3}
\end{table}

\begin{figure}[h]
\includegraphics[width=5.5cm, angle=-90]{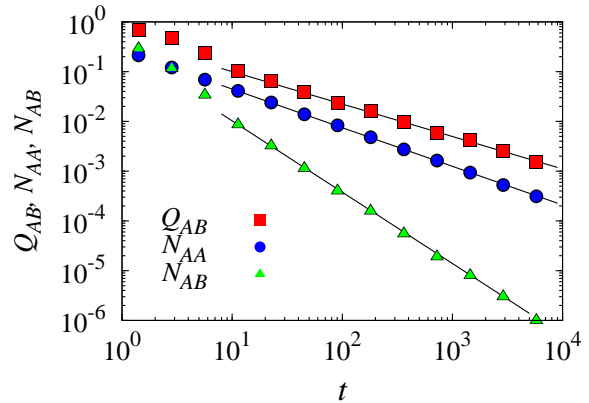}
\caption{(Color on line) Plot of $Q_{AB}$, $N_{AB}$ and $N_{AA}$ versus time $t$
for a (3,3)-flower network. The slopes of guidelines are $-0.65$, $-0.78$ and $-1.43$
from the top.}\label{q_ab}
\end{figure}

To see if the segregation of $A$-rich or $B$-rich domains forms,
we examine a quantity,
\begin{equation}
Q_{AB}(t)=\frac{N_{AB}}{N_{AA}+N_{BB}},
\end{equation}
where $N_{AB}(t)$ is the number of ($A,B$) pairs located at the
nearest neighbors averaged over different initial configurations.
$N_{AA}$ and $N_{BB}$ are similarly defined~\cite{gallos2007}.
If $Q_{AB} \to 0$, then there is few pairs of
different species at neighbor nodes, whereas if $Q_{AB} \to 1$,
particles are mixed randomly. Since the particle density decreases in time,
their separation becomes large and two particles hardly locate at the nearest neighbors.
We examine $N_{AB}$ and $N_{AA}$ independently as a function of time.
Interestingly, they decrease with time in a power-law manner as shown in
Fig.~\ref{q_ab}, which can be explained as follows:

First, we examine $N_{AA}$. The linear size $\ell_{d}$ of a domain containing
a species grows with time as $\sim t^{1/d_w}$. A typical closest distance
$\ell_{AA}$ between two particles of the same species scales
as $\sim (1/\rho)^{1/d_f}$. Assuming that $\rho(t)\sim t^{-d_s/4}$,
one can obtain that $\ell_{AA}\sim t^{1/(2d_w)}$~\cite{leyvraz}.
When $d_s \le 2$, the case of concern in this Letter, random walks are compact
within the diffusion volume $\ell_d^{d_f}$, and thus
that is also valid within the volume $\ell_{AA}^{d_f}$. The probability to find
two such particles at the nearest neighbors is $1/\ell_{AA}^{d_f}$.
Thus $N_{AA}$ scales as $(1/\ell_{AA}^{d_f})\rho(t)$. That is,
\begin{equation}
N_{AA}(t) \sim t^{-d_s/2}.
\end{equation}

Second, we examine $N_{AB}(t)$. When two particles of different species arrive
at the nearest neighbors in the diffusion process, they can annihilate at the next step
with a finite probability. Thus, we may set $N_{AB}(t)\propto {d\rho}/{dt}$,
and obtain that
\begin{equation}
N_{AB}(t) \sim t^{-d_s/4-1}.
\end{equation}

Next, $Q_{AB}$ is obtained as $N_{AB}/N_{AA}$. We compare the results obtained
from simple arguments with numerical ones in Table~\ref{table4}.

\begin{table}[h]
\caption{Comparison of the exponents for $N_{AA}$ and $N_{AB}$ between theoretical
and numerical values for various embedded spaces, one dimensional regular lattice (1 dim), two dimensional square lattice (2 dim), critical branching trees (CBT) with
$\gamma=2.5$ and $\gamma=4.0$, and (3,3)-flower hierarchical network.}
\begin{center}
\begin{tabular}{lll}
\hline\hline
~~~{space}~~~&~~~~~~~~~~~~~~~$N_{AA}$~~~~~~&~~~~~~~~~~~$N_{AB}$~~~~~~~~~~\\
\end{tabular}
\begin{tabular}{ccccc}
~~~&~$d_s/2$~&~{\rm Num.}~&~$(d_s/4)+1$~&~{\rm Num.}~~\\
\hline
~1 dim~&~0.5~&~0.49~&~~1.25~~&~~1.27~~~\\
~2 dim~&~1.0~&~0.99~&~~1.50~&~~1.59~~~\\
~CBT ($\gamma=2.5$)~&~0.75~&~0.59~&~~1.38~&~~1.28~~~\\
~CBT ($\gamma=4.0$)~&~0.67~&~~~0.63~~~&~~1.33~&~~1.28~~~\\
~(3,3)-flower~&~0.82~&~0.78~&~~1.41~&~~1.43~~~\\
\hline
\end{tabular}
\end{center}
\label{table4}
\end{table}

\begin{figure}[t]
\includegraphics[width=5.5cm, angle=-90]{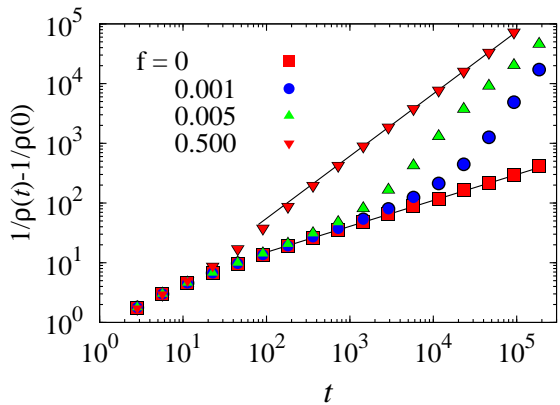}
\caption{(Color on line) The particle density versus time on rewired networks from a (3,3)-flower network. The slope increases as the fraction $f$ of rewired links increases from $f=0$ to $f=0.5$. For the $f=0$ and $f=0.5$ cases, the slopes are close to $\alpha\approx 0.43$ and 1, respectively. Since $\gamma\approx 3.58 > 3$, $\alpha=1$ is the mean-field result.}\label{rewire1}
\end{figure}

\begin{figure}[h]
\includegraphics[width=5.5cm, angle=-90]{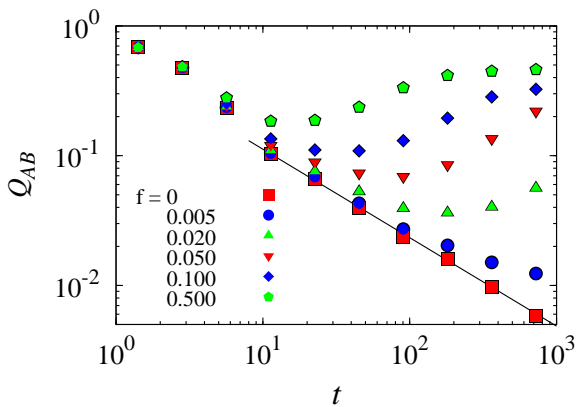}
\caption{(Color on line) Plot of $Q_{AB}$ as a function of $t$ for the rewired networks used in Fig.~\ref{rewire1}. Slope of the solid line is $-0.65$.}\label{rewire2}
\end{figure}

To confirm that the segregation is caused by local hubs in the fractal structures,
we destruct the local hubs by rewiring the links in the (3,3)-flower network
while conserving the degree distribution. Fig.~\ref{rewire1} shows that the exponent $\alpha$ changes from $\alpha \approx 0.43$ to the mean field value $\alpha=1$ as the number of rewired links increases. Moreover, $Q_{AB}$ does not decrease monotonically for the rewired
networks as shown in Fig.~\ref{rewire2}.

\begin{figure}[t!]
\includegraphics[width=5.5cm, angle=-90]{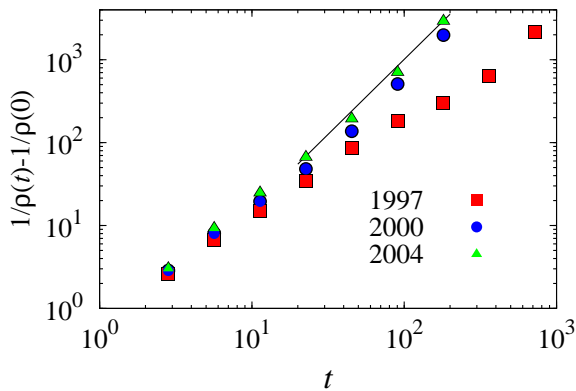}
\caption{(Color on line) The particle density as a function of time on the Internet in several different network profiles, the years 1997, 2000, and 2004. The slope of guideline is $1.8$.}\label{as}
\end{figure}

While many complex networks in real world are fractals, the Internet at the autonomous system level is not a fractal. This may be caused from the geographical effect. Due to this non-fractality, the segregation does not occur in the Internet in the two-species annihilation, and thus the particle density decreases fast with exponent $\alpha\approx 1.8$ from recent Internet topology in the year 2004 as shown in Fig.\ref{as}. This property can be used beneficially when one designs a protocol for P2P network, virus-antivirus annihilation robot, etc~\cite{syook}.

It is noteworthy that whereas the decaying behavior obeying the formula (\ref{rt_origin}) applies to the BA model when $m$, the number of incoming links at each time step, is larger than 1, it is not so for the BA tree network with $m=1$.
This is because the tree structure has limited paths, which enhances segregation.
Thus, the particle density decays slowly with exponent $\alpha\approx 0.5$, even
though the BA tree is not fractal.

In summary, the segregation effect in the two-species annihilation reaction dynamics has to be taken into account when the dynamics takes place on fractal, modular, or tree networks. In this case, the role of hubs is different from that of random SF networks and the particle density
decays slowly in a power-law manner with exponent less than 1, even though those networks are scale free.

This work was supported by KOSEF grant Acceleration Research (CNRC) (No.R17-2007-073-01001-0).

\end{document}